\documentclass[epjCONF]{svjour}
\usepackage{graphics}
\usepackage{epsfig}
\usepackage[varg]{txfonts} 
\usepackage[latin1]{inputenc}
\session-title{International Conference on Frontier Physics 2012}
\begin{document}
\title{Recent Results of the CMS Experiment}
\author{Tommaso Dorigo\inst{1}\fnmsep\thanks{\email{dorigo@pd.infn.it}} for the CMS Collaboration }
\institute{INFN - Sezione di Padova}
\abstract{
The CMS experiment obtained a large number of groundbreaking results from the analysis of 7- and 8-TeV proton-proton collisions produced so far by the Large Hadron Collider at CERN. In this brief summary only a sample of those results will be discussed. A new particle with mass $m_H = 125.3 \pm 0.4 (stat.) \pm 0.5 (syst.)$ GeV and characteristics compatible with those expected for a standard model Higgs boson has been observed in its decays to photon pairs, $WW$ pairs, and $ZZ$ pairs. Searches for the rare decays $B_d \to \mu \mu$ and $B_s \to \mu \mu$ have allowed to set limits on the branching fractions which are close to standard model predictions, strongly constraining new physics models. The top quark has been studied with great detail, obtaining among other results the world's best measurement of its mass as $m_t = 173.49 \pm 0.43 (stat.+JES) \pm 0.98(syst.)$ GeV. New physics models have been strongly constrained with the available data.  
} 
\maketitle
\section{Introduction}
\label{intro}

The observation of a particle with characteristics closely matching those predicted for a standard model Higgs boson, announced jointly by the ATLAS and CMS experiments on July $4^{th}$ this year, constitutes a turning point for the experiments at the CERN Large Hadron Collider (LHC). The main purpose of building the LHC and its giant detectors was indeed that of discovering the Higgs boson and understanding the origin of electroweak symmetry breaking: we have completed a very important step in that direction, but we are certainly only at the beginning of a long journey. Even assuming that the new 125 GeV resonance is indeed the standard model Higgs boson many questions remain unanswered; in the process of answering those questions we might however discover that the new-found particle is a Supersymmetric Higgs boson, or an even more exotic object. In other words, the study of the properties of the new 125 GeV particle constitutes a whole new chapter at the frontier of particle physics, one which will take the next decade to write.

In parallel with the new particle discovery, the CMS experiment has kept expanding our knowledge of frontier particle physics with several new precise measurements of electroweak physics observables, and with the search of new particles and phenomena predicted by physics models extending the standard model. In the present document we can only supply a couple of examples of the wide-range searches and measurements that are being produced from the analysis of 7- and 8-TeV proton-proton collision data.

Section~\ref{s:cms} describes briefly the experimental environment. Our summary of the recent CMS results starts in Sec.~\ref{s:SM}, where we highlight the precise new measurements of top quark mass and the search for rare $B$ meson decays. In Sec.~\ref{s:higgs} we briefly review the current measurements of Higgs boson mass, cross section, and properties. We briefly mention searches for new physics in Sec.~\ref{s:exotica}. 
We offer some conclusions in Sec.~\ref{s:conclusions}.

\section {The CMS Detector}
\label{s:cms}

\begin{figure}[h!]
\centerline{\epsfig{file=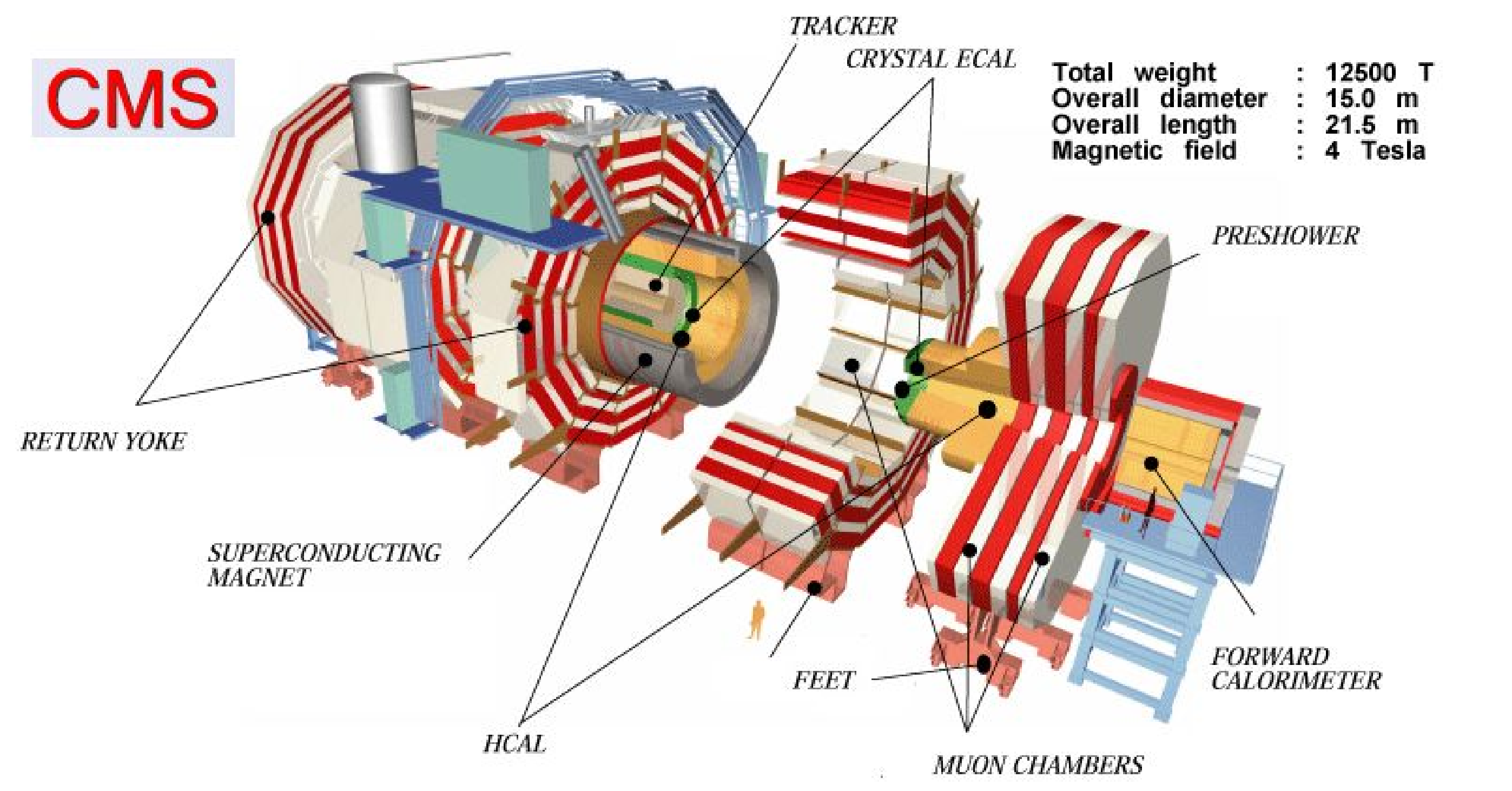,width=13cm}}
\caption{\em An exploded view of the CMS detector, showing the outer muon chambers  (white) embedded in iron (red). Internally can be seen the calorimeter system (ECAL, in green, and HCAL, in orange). The tracker is located in the core of the central barrel.}
\label{f:cms}
\end{figure}

CMS --an acronym for Compact Muon Solenoid-- is a multi-purpose magnetic detector designed to study proton-proton collisions delivered by the CERN Large Hadron Collider. The detector is located in a underground cavern at a depth of 100m at the site of Cessy, near the border of France and Switzerland. Particles emitted in hard collisions at the center of CMS cross in succession a silicon tracker, electromagnetic and hadron calorimeters, a solenoid magnet, and muon drift chambers embedded in the solenoid iron return yoke. A drawing of the CMS detector is shown in Fig.~\ref{f:cms}.

\subsection{Overview of the CMS Detector}

The momenta of charged particles emitted in the collisions at the center of the CMS detector are measured using a 13-layer silicon pixel and strip tracker; 66 million silicon pixels of dimensions 100x150 $\mu m$ are arranged in three barrel layers, and are surrounded by 9.6 million 180 $\mu m$-wide  silicon strips arranged in additional concentric barrels in the central region and disks in the endcap region. In order to allow a precise measurement of charged particle momenta, the silicon tracker is immersed in the 3.8 T axial field produced by a superconducting solenoid. The tracker covers the pseudorapidity range $|\eta| < 2.5$, where pseudorapidity is defined as $\eta = - ln \tan \theta /2$ and $\theta$ is the polar angle of the trajectory of a particle with respect to the direction of the counter-clockwise proton beam. 

	Surrounding the tracker are an electromagnetic calorimeter (ECAL) composed of lead tungstate crystals, and a brass-scintillator hadron calorimeter (HCAL). These detectors are used to measure the energy of incident particles from the produced electromagnetic and hadronic cascades; they consist of a barrel assembly covering the central region, plus two endcaps covering the solid angle for particles emitted at lower angle with respect to the beams direction. The ECAL and HCAL extend to a pseudorapidity range of $|\eta|< 3.0$; at still smaller angles particles emitted in the collision encounter a steel/quartz-fiber Cherenkov forward detector (HF) which extends the calorimetric coverage to $|\eta| < 5.0$. 

	The outermost component of the CMS detector is the muon system, consisting of four layers of gas detectors placed within the steel return yoke. The CMS muon system performs a high-purity identification of muon candidates and a stand-alone measurement of their momentum, and in combination with the inner tracker information provides a high-resolution determination of muon kinematics. More detail on the CMS detector is provided elsewhere~\cite{ref13}.
	
	CMS collects data with a two-level trigger system. Level 1 is a hardware trigger based on custom-made electronic processors that receive as input a coarse readout of the calorimeters and muon detectors and perform a preliminary selection of the most interesting events for data analysis, with an output rate of about 100 kHz. Level 2, also called "High-Level Trigger" (HLT), uses fine-grained information from all sub-detectors in the regions of interest identified by Level 1 to produce a final decision, selecting events at a rate of about 300 Hz  by means of speed-optimized software algorithms running on commercial computers.

\subsection{ The LHC in 2011 and 2012}

The 2011 proton-proton run of the LHC started on March $14^{th}$ and terminated on October $30^{th}$. Proton-proton collisions were produced at the centre-of-mass energy of 7 TeV. In the course of seven months of data taking CMS acquired a total of 5.3 inverse femtobarns of integrated luminosity; 5.0 of these were collected with all the CMS subdetectors fully operational. 

	During the 2011 run the instantaneous luminosity reached up to $3.5 \times 10^{33} cm^{-2} s^{-1}$. At a bunch crossing rate of 50 ns, the average number of pp interactions per bunch crossing was approximately 10. In such conditions, the rare hard collision which produces the physics objects (electrons, muons, taus, photons, energetic jets, missing transverse energy) recognized by the trigger system and fulfilling the criteria for data acquisition is usually accompanied by several additional pp interactions overlapping with it in the same bunch crossing. These additional collisions, which are typically of low energy but may still produce significant contributions to global event characteristics such as total visible energy or charged particle multiplicity, are denoted as pile-up events. The analysis of the hard collision properly includes the effect of pile-up, which is also modeled in all the necessary Monte Carlo (MC) simulated samples.

In 2012 the Large Hadron Collider has been operating at the increased energy of 4 TeV per beam, for a centre-of-mass energy of 8 TeV. The run started on April $4^{th}$, and at the time of writing (late September) over $15 fb^{-1}$ of proton-proton collisions have been delivered to the CMS experiment.

The LHC in 2012 has been running at higher instantaneous luminosities than in 2011. The peak instantaneous luminosity usually peaks at $7.5 \times 10^{33} cm^{-2} s^{-1}$. This produces a higher pileup than in 2011; CMS has responded to the resulting reconstruction challenge with more sophisticated algorithms and calibration procedures, which have allowed to maintain the same physics output despite the harsher experimental conditions. The results discussed in the following sections are based on data collected in 2011 and until June 2012, for a total of up to $10 fb^{-1}$.

\section{Precision Measurements of Electroweak Observables}
\label{s:SM}

At centre-of-mass energies of proton-proton collisions of 7 TeV and above, the LHC can be aptly described as a top quark factory. As an example, of the order of 800 thousand top quark pairs have been produced in the core of CMS in the 2011 run. Even larger is of course the number of produced W and Z bosons (respectively 500 millions and 150 millions). Electroweak interactions parameters can be determined with unprecedented accuracy by the analysis of CMS data, challenging theoretical predictions. In what follows we summarize only a few of the many new measurements produced by CMS with vector bosons and top quarks.

\subsection{ Vector Boson Production Cross Sections}

The production cross section of W and Z bosons at a centre-of-mass energy of 7 and 8 TeV has been studied both inclusively~\cite{ref14,v8tev} and as a function of the number of hadronic jets accompanying the bosons~\cite{ref15}. Additional measurements have determined the cross section of $W \gamma$ and $Z\gamma$ production~\cite{ref16}, as well as the production rate of WW, WZ, and ZZ pairs~\cite{ww36pb,ref17,ww8tev,zz8tev}. In all these measurements, W boson candidates have been selected by searching for their decay to $e \nu_e$ and $\mu \nu_\mu$ final states, and Z bosons by searching for their ee and $\mu \mu$ final states. In the case of ZZ production the second boson has also been identified in its decay to $\tau$-lepton pairs.
 
\begin{figure}[h!]
\epsfig{file=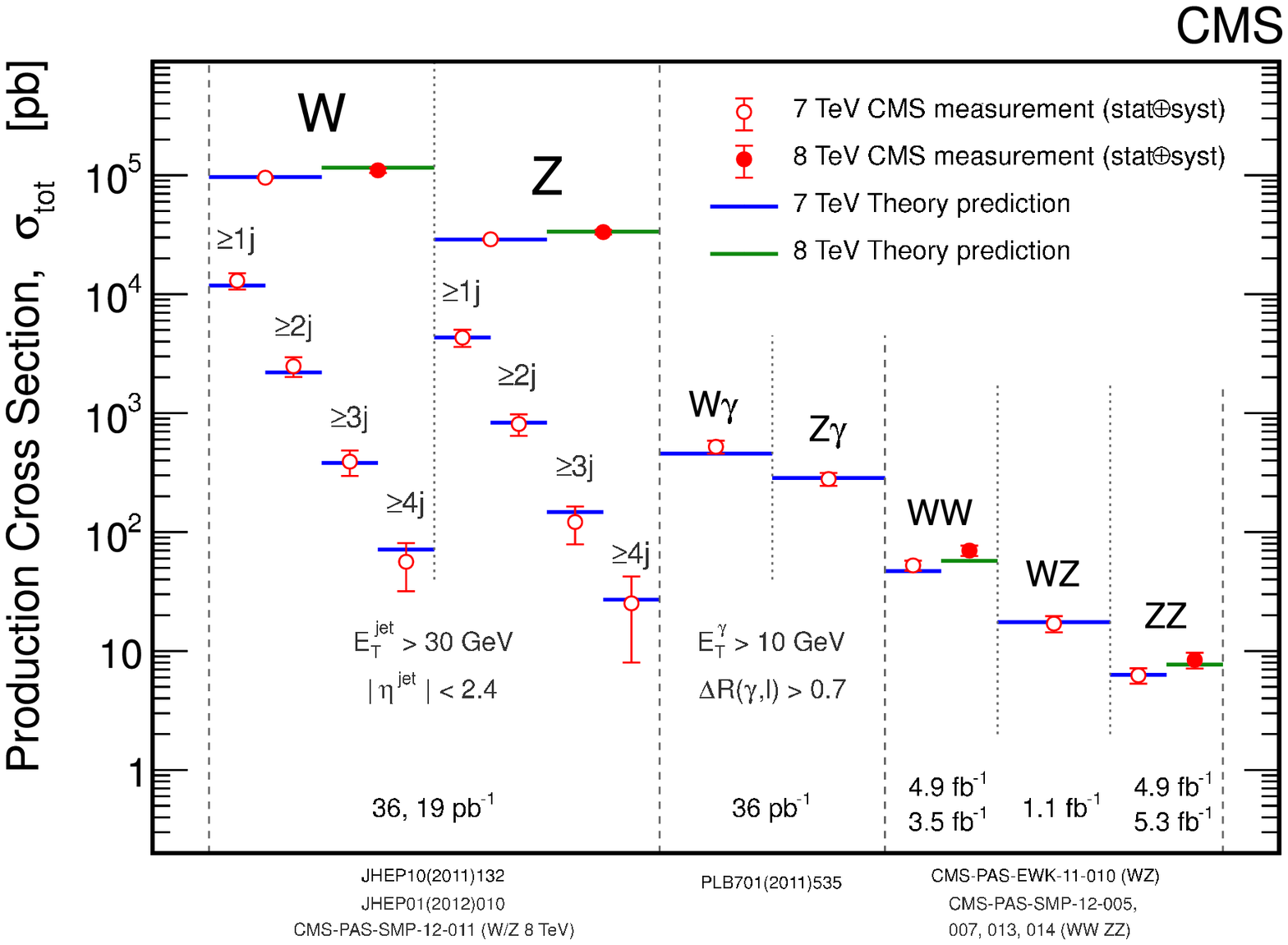,width=14cm}
\caption{\em The total production cross section of final states including W and Z bosons, in picobarns (red markers) are compared to theoretical predictions (blue lines). For single boson production are also reported the cross sections of processes including at least one to at least four hadronic jets with transverse energy $E_T>30$ GeV and pseudorapidity $|\eta|<2.4$.  }
\label{f:wzxs}
\end{figure}

The general picture that can be drawn is one of excellent agreement with theoretical calculations, which are available at next-to-leading order (NLO)~\cite{ref18,ref19,ref20} and next-to-next-to-leading order (NNLO)~\cite{ref21,ref22,ref23,ref24,ref25} in perturbative QCD. Figure~\ref{f:wzxs} provides a nice summary of the CMS measurements of these processes.

\subsection{ Top Quark Mass and Cross Section Measurements}

The large samples of top quark events produced in the 2011 run of the LHC have allowed the CMS experiment to measure with great accuracy the top pair production cross section in 7- and 8-TeV proton-proton collision using several different final states. The most precise CMS determinations come from the analysis of the clean dilepton final state of top pair decay~\cite{ttxs7,ttxs8}. These have reached an equal or smaller total uncertainty than existing theoretical estimates at NLO~\cite{ttxsnlo1} and NNLO~\cite{ref29,ref31}: with data corresponding to 2.3 inverse femtobarns the 7-TeV production cross section is measured at $\sigma_{t\bar{t}}^{7 \, TeV}=162 \pm 2 \pm 5 \pm 4 pb$, and with data corresponding to 2.4 inverse femtobarns of 2012 integrated luminosity the 8-TeV cross section is measured at $\sigma_{t \bar{t}}^{8 \, TeV} = 227 \pm 3 \pm 11 \pm 10 pb$. In both cases the first two quoted uncertainties are statistical and systematic, while the latter comes from the uncertainty in the total integrated luminosity.



\noindent
The top quark mass remains a parameter of great interest even after the precise measurements produced by the Tevatron experiments. The most recent CMS result~\cite{topmass}, which employs the full statistics of the 2011 run, is the most precise top quark mass determination in the world at the time of writing. The analysis uses top pair decays in the ``lepton plus jets'' topology, which arises when one of the two W bosons emitted in the $t \bar{t}\to W^+b W^- \bar{b}$  reaction decays via $W \to l \nu$, while the other W boson creates a pair of quarks. By selecting events from lepton-triggered data which contain a clean and isolated electron or muon candidate of $p_T>30$ GeV and four or more hadronic jets of $E_T>30$ GeV, two of them with identified secondary vertices from b-quark decay, a sample of 17,985 candidates is isolated. A kinematic fit using the world average value of the W boson mass as a constraint is used to determine an estimate of the top quark mass for each possible combination of jet assignments to the final state partons. The estimated masses of all combinations in 5174 events passing a selection based on the probability of the best fit are finally used in a global likelihood which combines information on the top quark mass and the jet energy scale factor, the latter obtained by comparing  the pre-constraint mass of the jet pair assigned to the W decay with the world average W mass. 

	The top quark mass is measured to be $m_t = 173.49 \pm 0.43 \pm 0.98$ GeV, where the first uncertainty quoted is the combination of statistical with jet-energy-scale-related systematic uncertainty, and  the second is the quadrature sum of all other systematic uncertainties.


\subsection{ Search for $B_d \to \mu \mu$ and $B_s \to \mu \mu$ decays}

In an era when heavy flavor physics was to be dominated by B factories -machines providing electron-positron collisions at the Y(4S) centre-of-mass energy- experiments studying heavy flavor properties in hadronic collisions have gone far beyond being complementary to the dedicated electron-positron experiments. In particular, the CDF experiment could match and in some cases surpass the precision of several BaBar and Belle results in the latter's own field of excellence, mostly thanks to its capability of triggering on the impact parameter of tracks, thus selecting large samples of hadronic $B$ and $D$ meson decays.
The triggering on track impact parameter, however, is not currently possible in CMS and ATLAS, mainly because of two otherwise strong points in the design of these experiments: the one-order-of-magnitude higher bunch-crossing rate of the LHC, together with the two-orders-of-magnitude larger number of readout channels of the silicon trackers of new generation of which the LHC experiments are endowed.
	
\begin{figure}[h!]
\centerline{\epsfig{file=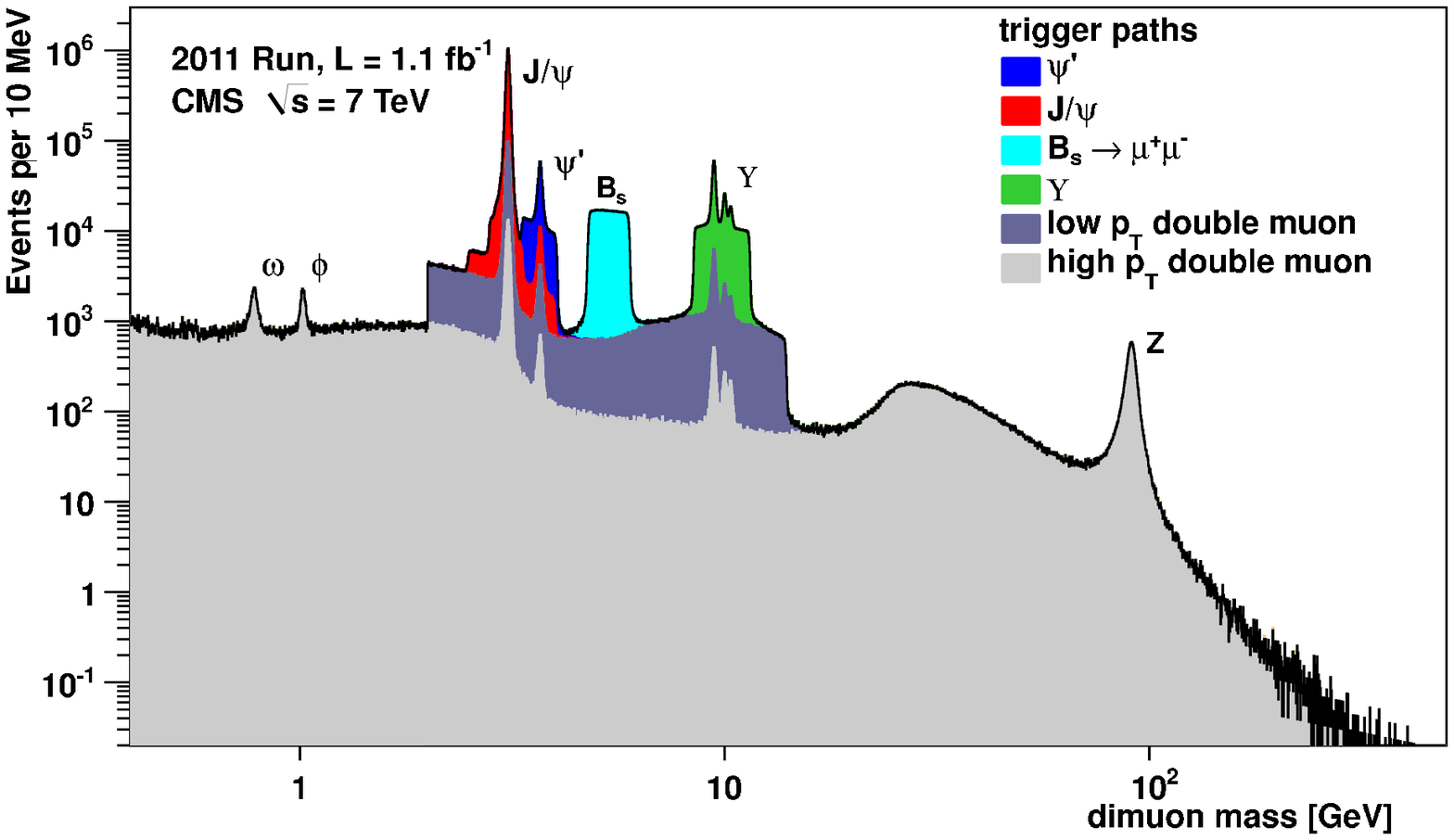,width=14cm}}
\caption{\em Invariant mass distribution of pairs of muon candidates of opposite sign collected by muon triggers in 1.1 inverse femtobarns of 7-TeV proton-proton collision in 2011. One immediately recognizes the signals due to muon pair decays of the  $\phi$, the $\omega$, the $J/\psi$ and the $\psi(2S)$, as well as the three lowest bottomonium states and the Z boson peak.}
\label{f:dimuon}
\end{figure}

The huge cross section of processes yielding bottom quarks in the final state of 7-TeV proton-proton collisions again comes to the rescue. CMS can collect semileptonic B-hadron decays of medium to large transverse momentum thanks to inclusive electron and muon triggers, as well as exploit the significant branching ratio of B-hadron decays to $J/\psi$ and $\psi(2S)$ mesons by triggering on the resulting pairs of low transverse momentum muons. Indeed, a simple graph showing the invariant mass of muon pairs collected by muon triggers speaks volumes about the heavy flavor potential of the experiment (see Fig.~\ref{f:dimuon}).

In the light of the above discussion, it is maybe not surprising that CMS has placed competitive limits on the rate of rare $B \to \mu \mu$ decays, nor that many other world-class results are being produced in this area of research. In the following we only make a brief mention of the search for the rare decays of neutral $B_d$ and $B_s$ mesons to muon pairs, which has been a hot topic in the last few years. These decays are heavily suppressed in the standard model due to the absence of flavor-changing neutral-current diagrams at tree level. Two additional factors further reduce their rate: the ratio $m^2/m_B^2$ between the squared masses of muons and B mesons implied by the helicity configuration of zero-total-momentum energetic fermion-antifermion final states, and the ratio $f_B^2/m_B^2$ (where $f_B$ is the B decay constant) due to the inner annihilation of quarks in the decaying meson. The smallness of the total predicted branching fractions, B($B_s \to \mu \mu$) = $(3.2 \pm 0.2)10^{-9}$  and $B(B_d \to \mu \mu) = (1.0 \pm 0.1) 10^{-10}$~\cite{ref36}, constitute an opportunity to search for indirect evidence of new physics, which could intervene in the form of the exchange of new virtual particles, with significant increases in the rate of these decays for specific values of the new physics parameters. 

	The search method is a counting experiment of events in the signal region of the dimuon mass distribution for both B species. Because of the dependence of mass resolution and background levels on the pseudorapidity of detected muons, two separate samples are analyzed independently and then combined: ``barrel'' candidates have both muons with $|\eta|<1.4$, and ``endcap'' candidates include all remaining events. MC simulations are used to estimate backgrounds from other B decays, while combinatorial backgrounds are evaluated from the data in suitable mass sidebands. A normalization sample of $B^+ \to J/\psi K^+$ decays, with the subsequent $J/\psi \to \mu \mu$ decay, is collected by a similar trigger to the one used for the rare decay search, and is used to remove the uncertainties of B hadron production cross section and integrated luminosity of the data sample.	


Muon candidates are selected using a variety of criteria, including the quality of their track fits. Muon pairs with invariant mass in the $4.9<M_{\mu \mu}<5.9$ GeV range are then preselected, and candidates in the signal regions ($5.2<M_{\mu \mu}<5.3$ GeV for the $B_d$ and $5.3<M_{\mu \mu}<5.45$ GeV for the $B_s$) are blinded to avoid potential bias in the selection procedure. A random-grid search of 1.6 million possible selection strategies is performed on a set of variables capable of discriminating the rare B decay signals from backgrounds, optimizing the sensitivity to the expected upper limit; among the used variables are the $\chi^2$ of the fit of muon trajectories to a common vertex, the transverse momentum of the higher-$p_T$ and the lower-$p_T$ muon, the B candidate transverse momentum, the isolation of the muons from other tracks in the event, the number of nearby tracks, and the smallest impact parameter of these tracks with respect to the common vertex of the muon pair. 


In the ``barrel'' sample two candidates are found for each B meson species, while in the ``endcap'' four $B_s$ candidates and zero $B_d$ candidates are observed. Upper limits on the branching ratios are determined at 95\% C.L. using the $CL_s$ criterion~\cite{ref37,ref38}: for the $B_s$ meson the limit is $B(B_s \to \mu \mu) < 7.7 \times 10^{-9}$, and for the $B_d$ meson the limit is $B(B_d \to \mu \mu) < 1.8 \times 10^{-9}$. These limits can be used to constrain several proposed extensions of the standard model. 

\clearpage
\section {Higgs Boson Physics} 
\label{s:higgs}

 On July $4^{th}$, 2012 the ATLAS and CMS collaborations released preliminary results of their Higgs boson searches. Both experiments claimed the observation of a new particle decaying into photon pairs or Z boson pairs, with a mass of approximately $m_H=125$ GeV; and both excluded basically all the remaining part of the searched mass spectrum from the LEP II lower limit of 114.4 GeV up to 600 GeV. The observed rates of decays of the new particle into the searched final states, as well as the production mechanisms inferred by the event topology, are in agreement with expectations for a standard model Higgs boson~\cite{ref54,ref55,ref56,ref57,ref58,ref59}. After a short review of the predicted phenomenology of Higgs production and decay,
we summarize below the status of the CMS measurements of Higgs boson mass, cross section, and properties. We refer the reader to the bibliography for preliminary papers describing the recent status of individual searches~\cite{hgg,hzz4l,hzz2l,hww,hwwlj,vhbb,ttbb,htt}.

\subsection{ Production and Decay}

The standard model Higgs boson has a non-zero coupling to all massive particles, and can therefore be produced by several different mechanisms in proton-proton collisions. The reactions studied so far at the LHC include gluon-fusion diagrams, where a Higgs boson is emitted most frequently by a virtual top-quark loop; vector-boson-fusion processes, where the Higgs is produced together with two characteristic high-rapidity hadronic jets resulting from the emission of two virtual W or Z bosons off the initial state quarks; and Higgs-strahlung diagrams where the particle is radiated by a highly-off-shell W or Z boson.


At a mass of $m_H=125$ GeV, the Higgs boson exhibits also a very rich decay phenomenology. The decays to weak boson pairs ($H \to WW^*$, $H \to ZZ^*$) are possible when one of the two final-state objects is off-mass-shell, but their branching ratios are comparatively small, allowing other processes to contribute significantly. The LHC experiments therefore search for five different decay modes of the Higgs boson: the two mentioned above as well as decays to b-quark pairs, $\tau$-lepton pairs, and photon pairs. The latter, although quite rare (with a predicted branching fraction of $2.3 \times 10^{-3}$), was in fact crucial for the first observation of the Higgs boson. Other still rarer decay modes ({\em e.g. $H \to Z \gamma$}) will also become accessible to investigation and measurement in the future.


\subsection{Determinations of Higgs boson cross section, mass, and coupling strengths}

CMS has recently published~\cite{higgsobservation} results of searches for the Higgs boson which employ the full 2011 statistics of proton-proton collisions and all data collected until June 2012. The following experimentally significant production modes of Higgs particles have been exploited: gluon-gluon fusion, vector-boson fusion, and Higgs-strahlung off vector bosons. Five decay modes have been considered: photon pairs~\cite{hgg}, Z boson pairs~\cite{hzz4l,hzz2l}, W boson pairs~\cite{hww,hwwlj}, bottom quark pairs~\cite{vhbb,ttbb}, and $\tau$-lepton pairs~\cite{htt}.

The results of these searches are combined by CMS by taking into account all statistical and systematic uncertainties and their correlations~\cite{ref76,ref70}. 
The combination is performed by constructing a global likelihood function, with each systematic source assigned to a nuisance parameter; each of these has a corresponding probability density function. Most of the systematic uncertainties are constrained by subsidiary measurements.

\begin{figure}[h!]
\begin{minipage}{0.49\linewidth}
\centerline{\epsfig{file=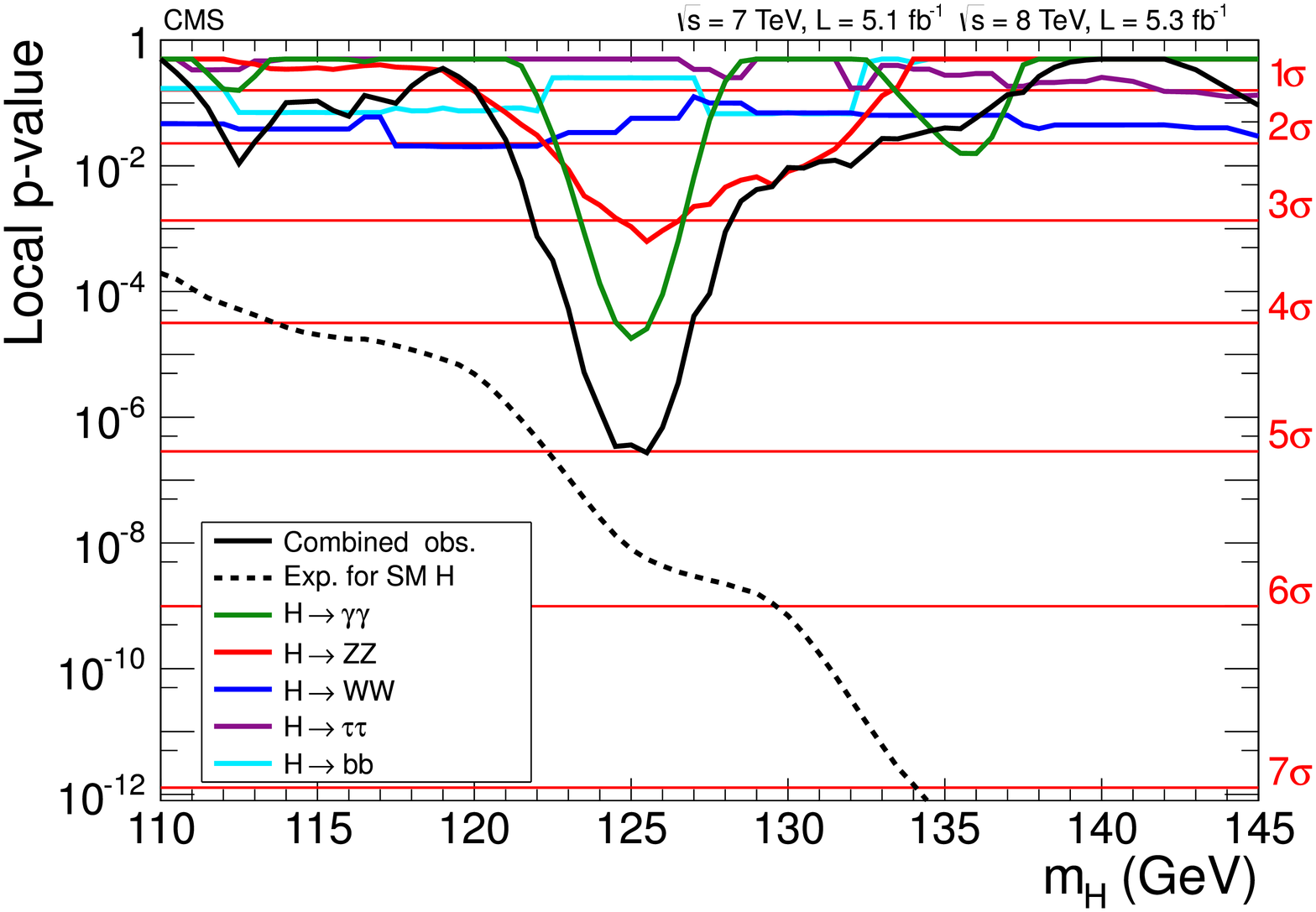,width=6cm}}
\end{minipage}  
\begin{minipage}{0.49\linewidth}
\centerline{\epsfig{file=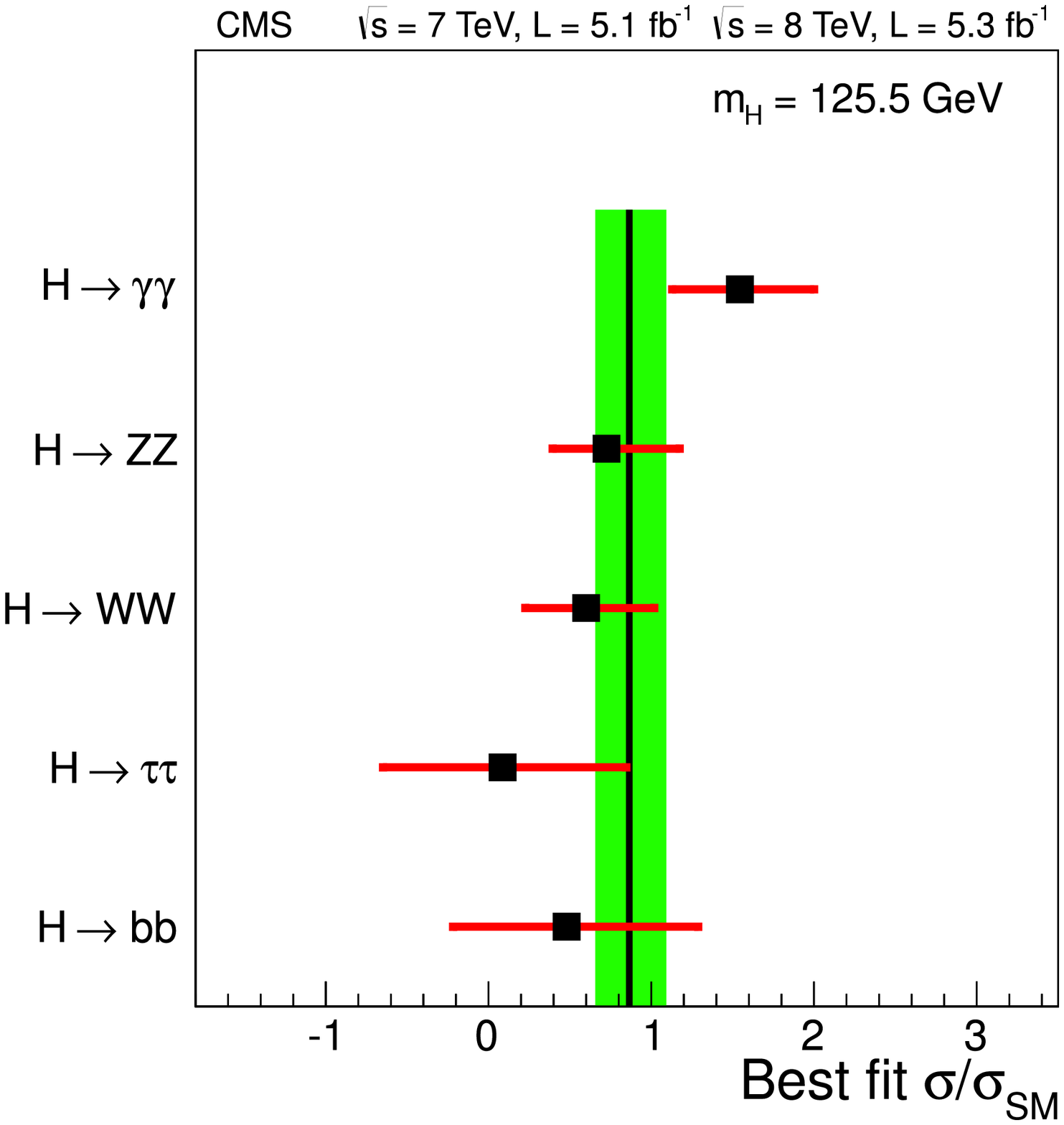,width=6cm}}
\end{minipage}  
\caption{\em Left: Local p-value of the background-only hypothesis as a function of $m_H$. Different search channels are indicated by different colored curves; the p-value predicted in the presence of a real Higgs boson is indicated with a dashed curve. Right:  Measured Higgs production cross sections for the studied decay modes, in units of the standard model prediction (a mass $m_H=125.5$ GeV is assumed). The shaded band shows the fitted average of experimental measurements.}
\label{f:pvalue_xs}
\end{figure}

The left panel in Fig.~\ref{f:pvalue_xs} shows the local p-value of observed excesses in the data, as a function of $m_H$. For $m_H=125$ GeV an excess corresponding to a $5 \sigma$ significance is observed, using the $\gamma \gamma$ and $ZZ$ final states. 

The different analyses provide rate measurements of Higgs decays in the various considered final states. The individual measurements are shown in the right panel of Fig.~\ref{f:pvalue_xs}, where the cross section is measured in units of the standard model expectation. The combined measurement of CMS is $\mu = 0.87 \pm 0.23$, in agremeent with the prediction.

\begin{figure}[h!]  
\begin{minipage}{0.49\linewidth}
\centerline{\epsfig{file=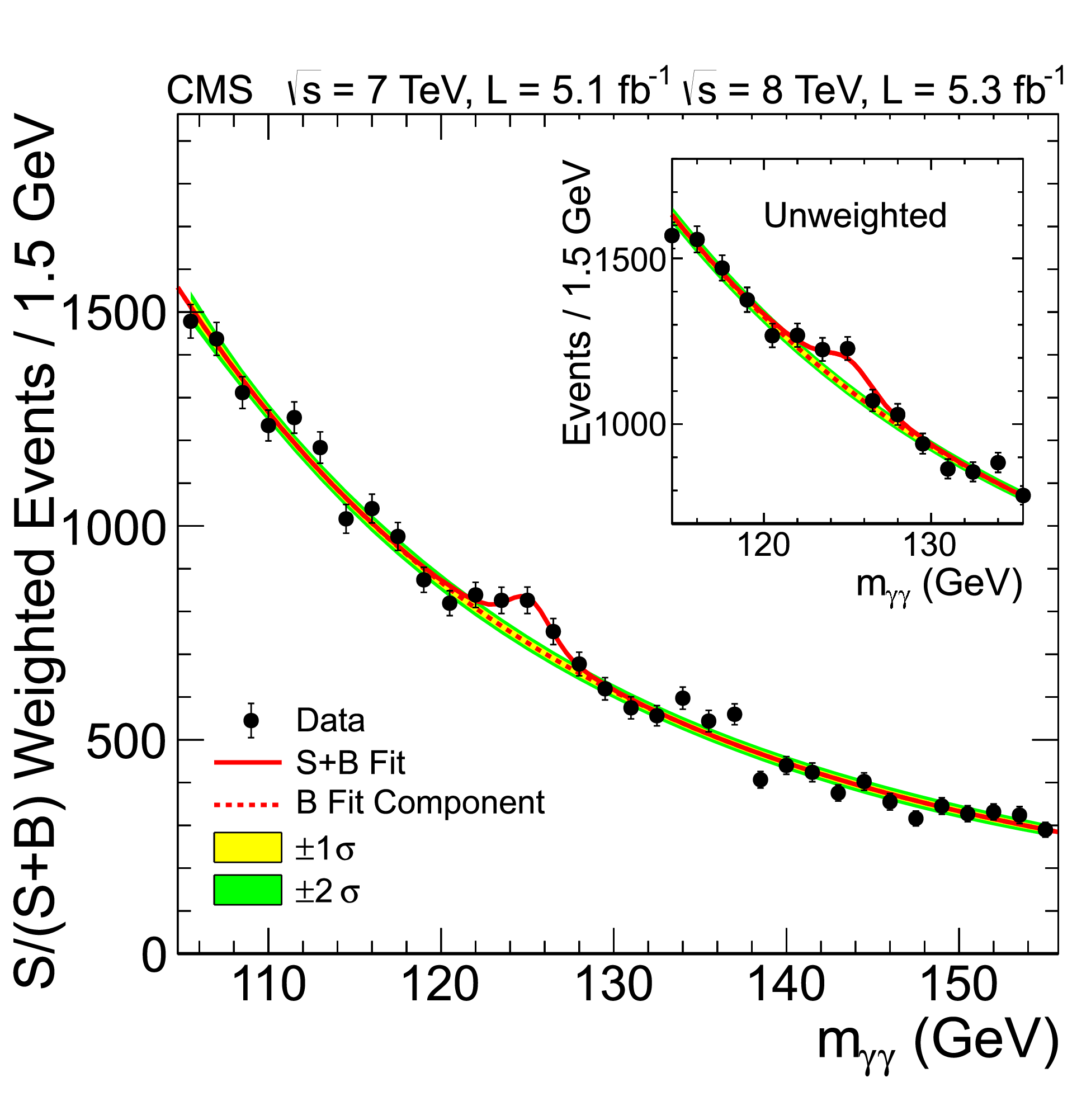,width=6cm}}
\end{minipage}  
\begin{minipage}{0.49\linewidth}
\centerline{\epsfig{file=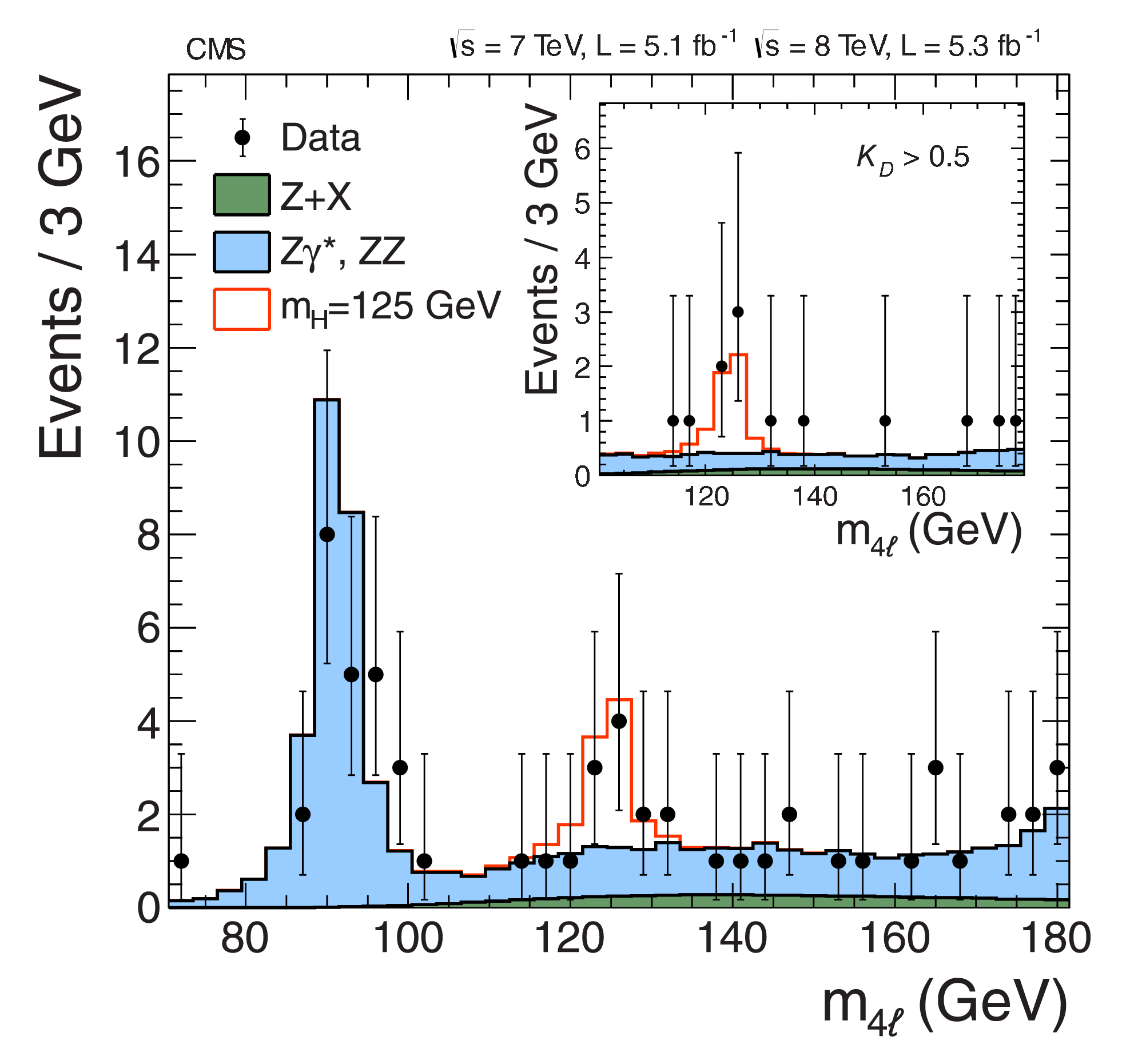,width=6cm}}
\end{minipage}  
\caption{\em Left: Reconstructed mass of photon pairs in $10.4 fb^{-1}$ of 2011+2012 data. Entries in the histogram in the main panel are weighted by signal-to-noise ratio. Right: Reconstructed mass of Z pairs. The tall peak on the left is due to $Z \to ll \gamma \to llll$ events; the inset on the upper right shows the result of a tighter selection based on a kinematic multivariate discriminant.}
\label{f:mass}
\end{figure}

\begin{figure}[h!]  
\begin{minipage}{0.49\linewidth}
\centerline{\epsfig{file=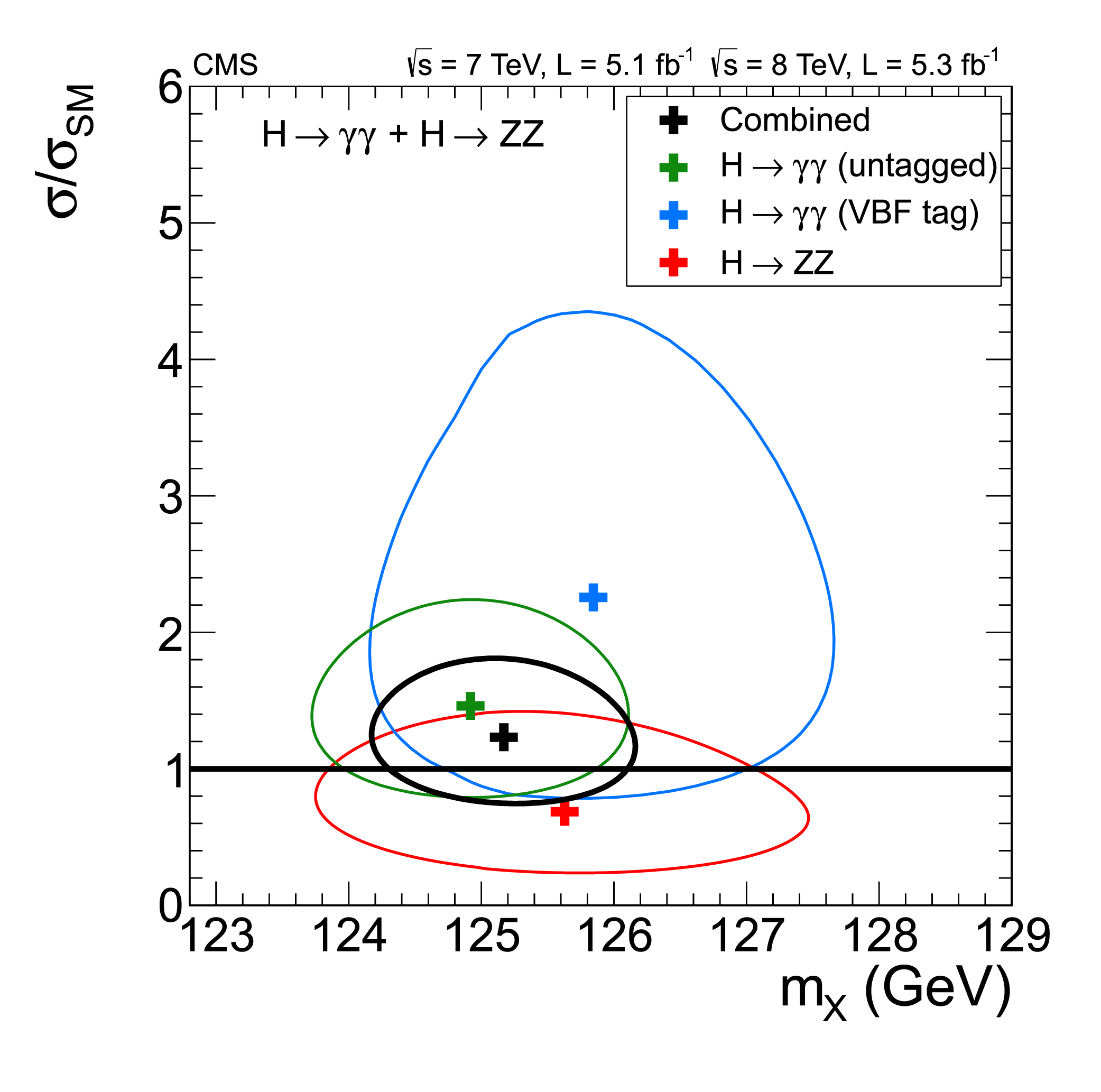,width=6cm}}
\end{minipage}  
\begin{minipage}{0.49\linewidth}
\centerline{\epsfig{file=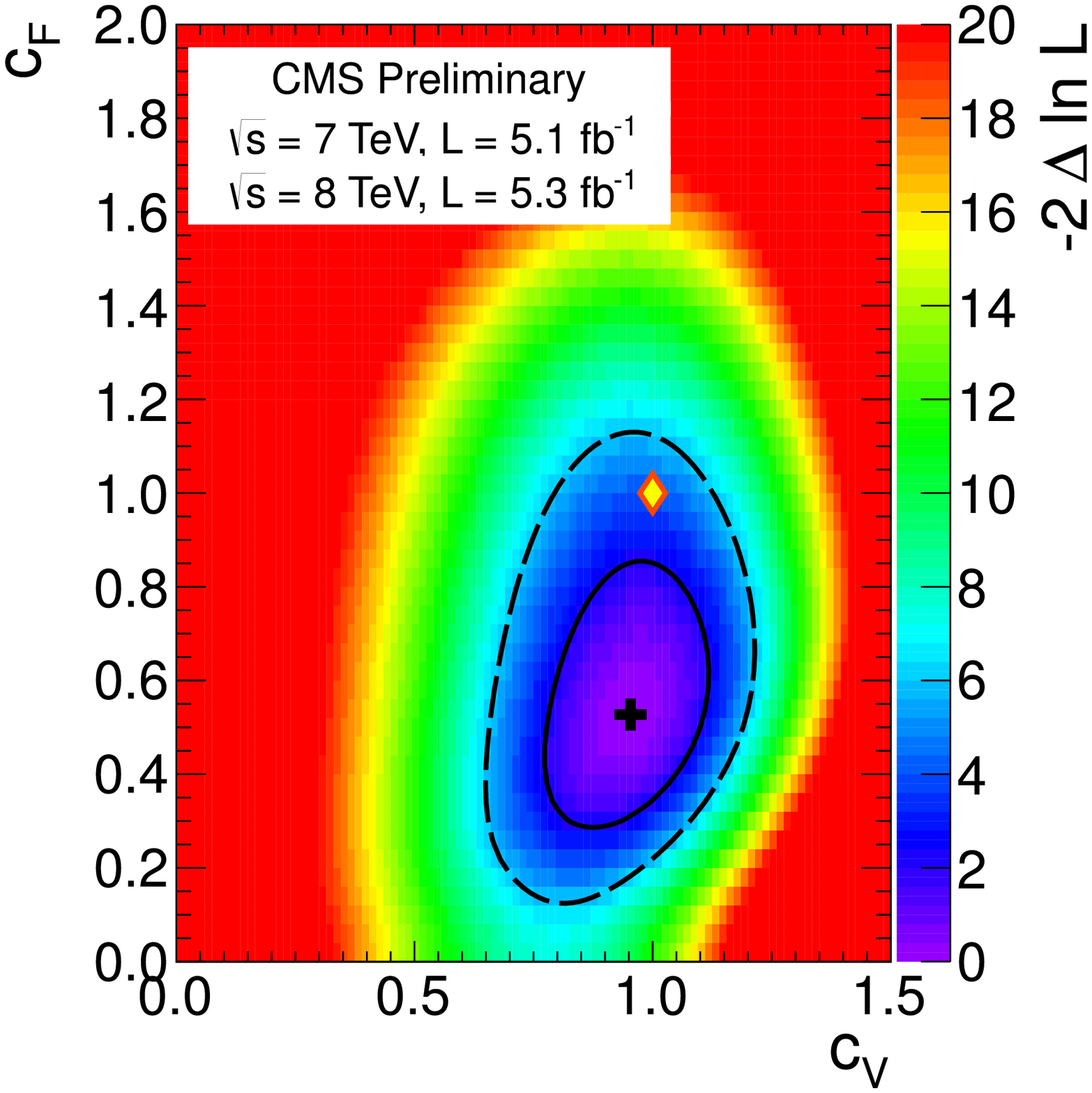,width=6cm}}
\end{minipage}  
\caption{\em Left: 1-sigma contours of mass and cross section measurements in the three bosonic final states, and their combination (black curve). Right: Fit results for the bosonic and fermionic coupling strength modifiers. The light-coloured marker at (1,1) indicates the standard model value.}
\label{f:properties}
\end{figure}

From the analysis of the diphoton and four-lepton final states, which are the ones with the largest signal-to-noise ratio and the best mass resolution, the Higgs mass has been measured to be $m_H = 125.3 \pm 0.4 \pm 0.5$ GeV, where the first uncertainty is statistical and the second one is systematic. Figure~\ref{f:mass} shows the signals in the reconstructed invariant mass distribution of diphoton and ZZ candidates. The left panel of Fig.~\ref{f:properties} shows the individual measurements in the mass-cross section plane, and their combination.

Finally, the observed signals can be used to fit for coupling strength modifiers, to allow for different coupling of the Higgs boson to fermions and bosons from the predictions of the standard model. The result is shown in the right panel of Fig.~\ref{f:properties}. Although the larger-than-expected rate of decays to photon pairs could be suggestive of anomalous couplings, the CMS measurement supports the standard model interpretation.

\section {Searches  \label{s:exotica}}

\subsection{ Searches for Supersymmetry Signatures}

Among all hypothesized extensions of the standard model, Supersymmetry (SUSY) is one of the most studied alternatives. The symmetry between standard model fermions and bosons with supersymmetric counterparts of bosonic and fermionic nature automatically cancels the large quantum contributions to the Higgs boson mass due to virtual loops of standard model particles~\cite{ref77,ref78,ref79,ref80}, solving the naturalness puzzle in a very elegant way; the added bonus is a unification-ready merging of coupling constants below the Planck scale. Supersymmetric particle searches have been carried out in the past thirty years without success, pushing the allowed mass of the hypothetical SUSY particles to higher and higher values. Despite those early results, the much larger centre-of-mass energy of the LHC led many to trust that SUSY particles would suddenly pop up soon after the start of data taking, with unmistakable and striking signatures. But Nature has chosen otherwise.

\begin{figure}[h!]
\centerline{\epsfig{file=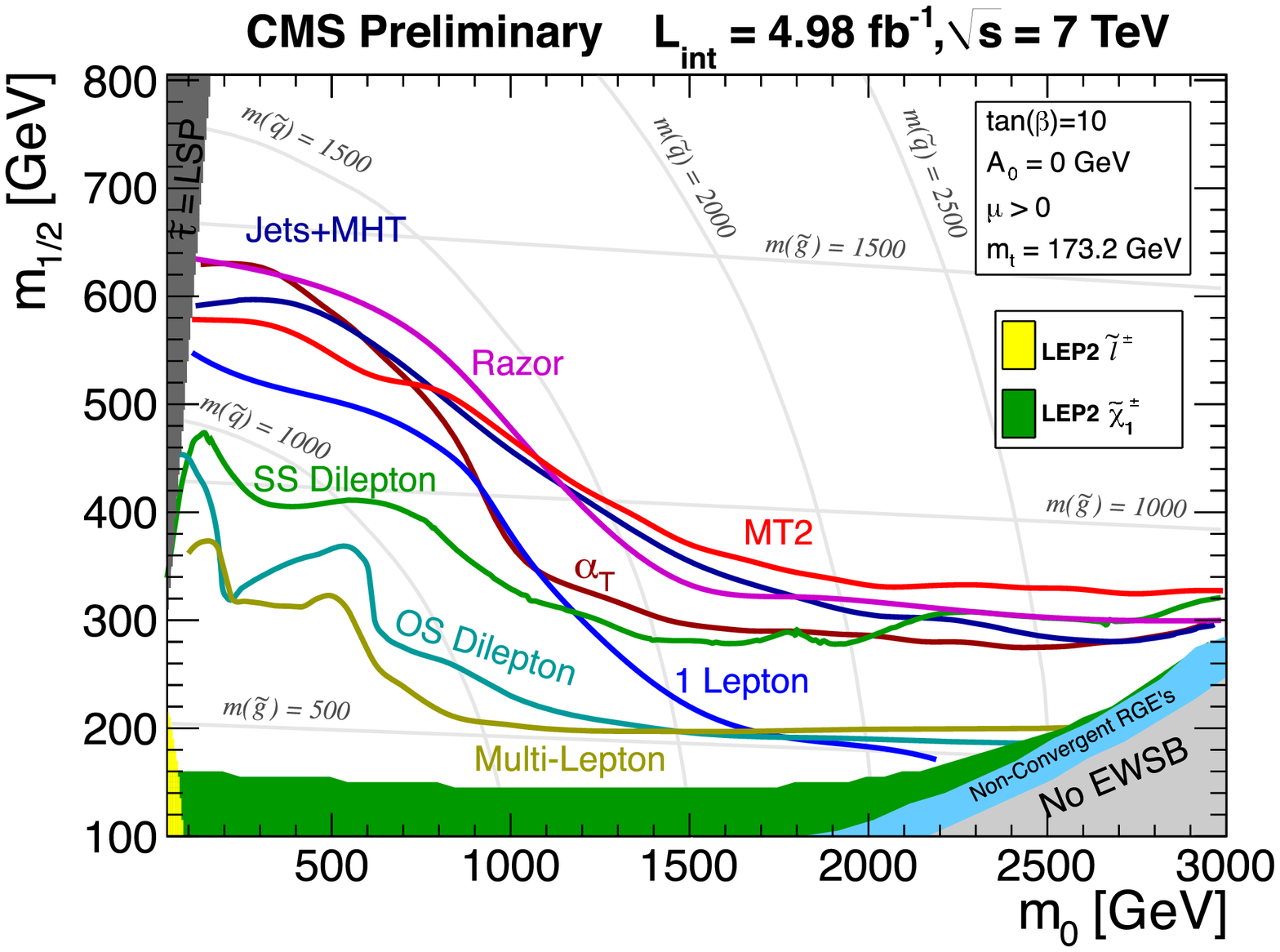,width=13cm}}
\caption{Summary of results of SUSY searches by CMS using 2011 data, here shown for a representative choice of SUSY parameters (see legend in the upper right corner). Grey curves show iso-contours in the value of squark and gluino masses.}
\label{f:susy}
\end{figure}

	The CMS experiment has searched the 2011 data for supersymmetric particle signatures in a number of final states and with a variety of advanced methods~\cite{nsunp1,nsunp2,nsunp3,newsus1,newsus2,newsus3,newsus4,ref81,ref82,ref83,ref84,ref85,ref86,ref87,ref89,ref90,ref91}. Here, for the sake of brevity, only a summary of those searches is provided. In general, SUSY particles can be copiously produced in LHC proton-proton collisions in the form of pairs of squarks or gluinos, which carry color quantum numbers and are thus subject to strong interactions. Depending on the mass spectrum of SUSY particles, the decay of squarks and gluinos may give rise to several lighter supersymmetric states in succession, with a typical ``cascade'' signature and characteristic kinematic features. At the end of the decay chain, a quite general signature of R-parity-conserving SUSY theories is the production of neutral weakly-interacting particles called neutralinos, which are the lightest in the supersymmetric spectrum and are thus stable. Their escape from the detector with large transverse momentum can be flagged by the same experimental observable used to detect neutrinos, i.e. a large energy imbalance in the plane transverse to the beams. Experimental searches often require large values of missing transverse energy, in some cases along with hadronic jets, in others accompanied with charged leptons or more complex final states.

	Figure~\ref{f:susy} summarizes the status of CMS searches for SUSY particles in 2011 datasets corresponding to up to 5 inverse femtobarns of integrated luminosity. Upper limits in production cross sections are turned into exclusion regions in the plane described by the universal scalar and gaugino masses. The most sensitive searches are the one for jets and missing energy and the "razor" analysis, which exploits the kinematic configurations of the jets in the reconstructed reference frame of super-particle decay. Those searches are expected to produce much tighter limits on superparticle masses when performed on the data from the full 2012 run, because of the larger statistics as well as the significant increase of the production cross section for very massive objects.

\subsection{ Other Exotica Searches }

A number of exotic extensions of the standard model have been tested by CMS; in no case a departure from standard model predictions has been observed. For lack of space here we may only refer the interested reader to the public pages of the CMS experiment~\cite{ref92} for a comprehensive list of all results of exotica searches.




\section{Conclusions}
\label{s:conclusions}

The CMS experiment has exploited the proton-proton collision data collected so far by the 2011 and 2012 runs of the Large Hadron Collider to produce a large number of groundbreaking results in precision measurements of standard model observables and searches for new physics. Among the most exciting of these results is certainly the observation of a new particle in the search for the Higgs boson~\cite{higgsobservation}; the particle has a cross section compatible with that expected from a standard model Higgs, decays that imply its bosonic nature, and a mass of $M_h=125.3 \pm 0.4 (stat.) \pm 0.5 (syst.)$ GeV; it is expected that the full 2012 dataset will allow more definite conclusions on the nature of this new boson. Another striking conclusion one can draw from the set of produced results is that natural low-scale Supersymmetry is getting close to be excluded across the board of the wide SUSY parameter space; similarly, other exotics new physics models are nowhere to be seen in TeV-scale collisions.

\clearpage
\section{ Acknowledgements}

We wish to congratulate our colleagues in the CERN accelerator departments for the excellent performance of the LHC machine. We thank the technical and administrative staff at CERN and other CMS institutes, and acknowledge support from: FMSR (Austria); FNRS and FWO (Belgium); Cap, CAPES, FAPERJ, and FAPESP (Brazil); MES (Bulgaria); CERN; CAS, Most, and NSFC (China); COLCIENCIAS (Colombia); MSES (Croatia); RPF (Cyprus); More, SF0690030s09 and ERDF (Estonia); Academy of Finland, MEC, and HIP (Finland); CEA and CNRS/IN2P3 (France); BMBF, DFG, and HGF (Germany); GSRT (Greece); OTKA and NKTH (Hungary); DAE and DST (India); IPM (Iran); SFI (Ireland); INFN (Italy); NRF and WCU (Korea); LAS (Lithuania); CINVESTAV, CONACYT, SEP, and UASLPFAI (Mexico); MSI (New Zealand); PAEC (Pakistan); MSHE and NSC (Poland); FCT (Portugal); JINR (Armenia, Belarus, Georgia, Ukraine, Uzbekistan); MON, Rosa tom, RAS and RFBR (Russia); MSTD (Serbia); MICINN and CPAN (Spain); Swiss Funding Agencies (Switzerland); NSC (Taipei); TUBITAK and TAEK (Turkey); STFC (United Kingdom); DOE and NSF (USA).

%
%

\end{document}